\RequirePackage{fix-cm}
\documentclass[reqno]{amsproc}
\usepackage{amsmath,amsfonts}
\usepackage{fixltx2e}
\usepackage{pstricks,pst-node,pst-plot}
\usepackage{graphicx}

\usepackage[nobysame]{amsrefs}
\usepackage{xyzbib}

\usepackage{mathtools}
    \mathtoolsset{showonlyrefs}

\usepackage{hyperref}

\newcommand{\F}[4]{%
    {}\,{}_2F_1\left(%
    \genfrac{}{}{0pt}{}{#1,\,#2}{#3}
    \bigg\vert#4\right)}

\newcommand{\rmi}{\mathrm{i}}
\newcommand{\rmd}{\mathrm{d}}

\begin{document}

\title{The limit shape of large alternating sign matrices}

\author{F. Colomo}
\address{INFN, Sezione di Firenze\\
Via G. Sansone 1, 50019 Sesto Fiorentino (FI), Italy}
\email{colomo@fi.infn.it}

\author{A.G. Pronko}
\address{Saint Petersburg Department of V.A.~Steklov Mathematical
Institute of Russian Academy of Sciences\\
Fontanka 27, 191023 Saint Petersburg, Russia}
\email{agp@pdmi.ras.ru}

\begin{abstract}
The problem of the  limit shape of large alternating sign matrices (ASMs) is
addressed by studying the emptiness formation probability (EFP)
in the domain-wall six-vertex model.
Assuming that the limit shape arises in correspondence to the
`condensation' of almost all solutions of the saddle-point
equations for certain multiple integral representation for EFP,
a conjectural expression for the limit shape of large ASMs is derived.
The case of $3$-enumerated ASMs is also considered.

\end{abstract}

\maketitle

\section{Introduction}

An alternating sign matrix (ASM) is a matrix of $1$'s, $0$'s and $-1$'s
such that in each row and column, all  nonzero entries alternate
in sign, and the first and the last nonzero entries are $1$'s.
In a weighted enumeration, or  $q$-enumeration,
ASMs are counted with a weight $q^k$, where $k$ is the total number of
$-1$'s in each matrix.
There are many nice results concerning ASMs, mainly devoted to their
various enumerations; for a review see, e.g., book \cite{Br-99}.

In this paper, we address the problem of the limit shape of large ASMs.
The problem comes out from the fact that in ASMs their corner regions mostly
contain $0$'s while in the interior there are many nonzero entries.
As the size of ASMs increases, the probabilities of finding $1$'s and $-1$'s in their entries
in the corner regions vanish, while in the central
region these probabilities remain finite.
When considering very large ASMs in an appropriate
scaling limit, (e.g., when large matrices are scaled to a unit square),
one can expect that such regions become sharply separated.
Assuming that the very fact of this phase separation is somehow ascertained,
an interesting question which can be then addressed concerns finding an explicit equation
for the spatial curve separating these two regions \cite{P-01}.

In essence, this curve can be regarded as an Arctic curve
for ASMs, similar to the Arctic Circle of domino tilings
of large Aztec diamonds \cite{JPS-98}. Since the Arctic curve
determines the shape of the internal, or `temperate', region of ASMs,
one can refer to this curve as the limit shape of large ASMs (cf.\/ \cite{CLP-98}).
More generally, one can address the same problem for
the case of $q$-enumerated ASMs.
Then the case $q=2$ corresponds to the Arctic Circle of the
domino tilings (see equation \eqref{ASM2eq} below).
The main result of the present paper consists in providing explicit expressions,
albeit conjectural, for the limit shapes of $1$- and $3$-enumerated ASMs.
They are given by equations \eqref{ASMeq} and \eqref{ASM3eq}, respectively.

To treat the problem, we exploit the one-to-one correspondence between
ASMs and configurations of the six-vertex model, which has been found and
efficiently used in papers \cite{EKLP-92,Ku-96,Ze-96}.
This correspondence takes place
when the six-vertex model is considered with the so-called domain wall
boundary conditions (DWBC). The six-vertex model with these boundary
conditions has been originally introduced and studied in
\cite{K-82,I-87,ICK-92}.

\begin{figure}
\centering
\addtolength{\arraycolsep}{-2pt}

\psset{unit=15pt}
\newcommand{\arr}{\lput{:U}{\begin{pspicture}(0,0)
\psline(0,.15)(.2,0)(0,-.15) \end{pspicture}}}

\newpsobject{ed}{pcline}{}

\begin{pspicture}(14,6)
\multirput(1,0)(1,0){5}{\ed(0,0)(0,1)\arr \ed(0,6)(0,5)\arr}
\multirput(0,1)(0,1){5}{\ed(1,0)(0,0)\arr \ed(5,0)(6,0)\arr}
\ed(2,5)(1,5)\arr\ed(3,5)(2,5)\arr\ed(3,5)(4,5)\arr\ed(4,5)(5,5)\arr
\ed(2,4)(1,4)\arr\ed(2,4)(3,4)\arr\ed(4,4)(3,4)\arr\ed(4,4)(5,4)\arr
\ed(1,3)(2,3)\arr\ed(3,3)(2,3)\arr\ed(3,3)(4,3)\arr\ed(4,3)(5,3)\arr
\ed(2,2)(1,2)\arr\ed(2,2)(3,2)\arr\ed(3,2)(4,2)\arr\ed(5,2)(4,2)\arr
\ed(2,1)(1,1)\arr\ed(3,1)(2,1)\arr\ed(4,1)(3,1)\arr\ed(4,1)(5,1)\arr
\ed(1,5)(1,4)\arr\ed(2,5)(2,4)\arr\ed(3,4)(3,5)\arr\ed(4,5)(4,4)\arr\ed(5,5)(5,4)\arr
\ed(1,4)(1,3)\arr\ed(2,3)(2,4)\arr\ed(3,4)(3,3)\arr\ed(4,3)(4,4)\arr\ed(5,4)(5,3)\arr
\ed(1,2)(1,3)\arr\ed(2,3)(2,2)\arr\ed(3,2)(3,3)\arr\ed(4,2)(4,3)\arr\ed(5,3)(5,2)\arr
\ed(1,1)(1,2)\arr\ed(2,1)(2,2)\arr\ed(3,1)(3,2)\arr\ed(4,2)(4,1)\arr\ed(5,1)(5,2)\arr
\rput(11,3){
$
\begin{pmatrix}
0 & 0 & 1 & 0 & 0 \\[3pt]
0 & 1 & -1 & 1 & 0 \\[3pt]
1 & -1 & 1 & 0 & 0 \\[3pt]
0 & 1 &  0& -1 & 1 \\[3pt]
0 & 0 & 0 & 1 & 0
\end{pmatrix}
$}
\end{pspicture}
\addtolength{\arraycolsep}{2pt}
\caption{A possible configuration of the six-vertex model with DWBC and
the corresponding
ASM, for $N=5$.}
\label{fig-dwbcasms}
\end{figure}
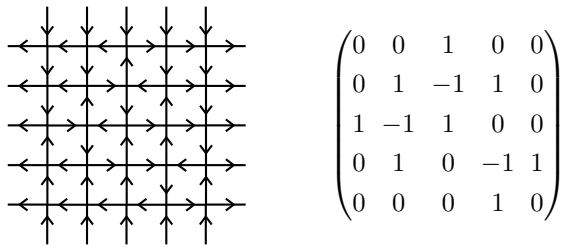

Using the standard description of local states in terms of arrows
(see, e.g., \cite{B-82}), DWBC mean that the model is considered on a square lattice of
$N$ vertical and $N$ horizontal lines, where the arrows
on all external horizontal edges point outward, while the arrows on all
external vertical edges point inward. Then, the $0$'s in ASM's entries
correspond to the vertices of weights $a$ and $b$, while both  $1$'s and
$-1$'s correspond to vertices of weight $c$, as shown in
Figure~\ref{fig-vertices}. Figure~\ref{fig-dwbcasms} shows a possible
configuration of the six-vertex model DWBC
and the corresponding ASM.

\begin{figure}
\centering
\psset{unit=15pt}
\newcommand{\arr}{\lput{:U}{\begin{pspicture}(0,0)
\psline(0,0.15)(.2,0)(0,-.15) \end{pspicture}}}


\begin{pspicture}(0,-1)(17,4)
\pcline(0,3)(1,3)\arr \pcline(1,3)(2,3)\arr
\pcline(1,2)(1,3)\arr \pcline(1,3)(1,4)\arr
\rput[B](1,.5){$a$}
\rput[B](1,-1){$0$}
\pcline(4,3)(3,3)\arr \pcline(5,3)(4,3)\arr
\pcline(4,4)(4,3)\arr \pcline(4,3)(4,2)\arr
\rput[B](4,.5){$a$}
\rput[B](4,-1){$0$}
\pcline(6,3)(7,3)\arr \pcline(7,3)(8,3)\arr
\pcline(7,4)(7,3)\arr \pcline(7,3)(7,2)\arr
\rput[B](7,.5){$b$}
\rput[B](7,-1){$0$}
\pcline(11,3)(10,3)\arr \pcline(10,3)(9,3)\arr
\pcline(10,2)(10,3)\arr \pcline(10,3)(10,4)\arr
\rput[B](10,.5){$b$}
\rput[B](10,-1){$0$}
\pcline(12,3)(13,3)\arr \pcline(14,3)(13,3)\arr
\pcline(13,3)(13,2)\arr \pcline(13,3)(13,4)\arr
\rput[B](13,.5){$c$}
\rput[B](13,-1){${-}1$}
\pcline(16,3)(17,3)\arr \pcline(16,3)(15,3)\arr
\pcline(16,4)(16,3)\arr \pcline(16,2)(16,3)\arr
\rput[B](16,.5){$c$}
\rput[B](16,-1){$1$}
\end{pspicture}
\caption{The six allowed arrow configurations, their weights,
and the corresponding entries of ASMs.}
\label{fig-vertices}
\end{figure}
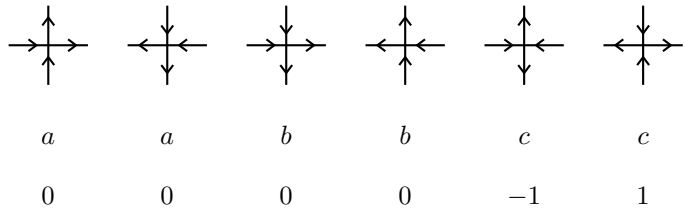

To count ASMs, the weights $a$, $b$, $c$ are to be put all to the same value,
e.g., $a=b=c=1$. In $q$-enumeration, ASMs are counted by taking
$a=b=1$ and $c=\sqrt{q}$, since in each configuration the
vertices of type five and six come in pairs, in addition to $N$
vertices of type six being always present due to DWBC.

The main tool which we use to address the problem of the Arctic curve of the domain-wall
six-vertex model is a particular, non-local, correlation function, the
so-called emptiness formation probability (EFP). In paper \cite{CP-07b}
a multiple integral representation for EFP has been derived. Here, we
study the multiple integral representation for large size of the
lattice (corresponding to large size of ASMs).

In our approach  an important role is played by the conjecture that
the Arctic curve appears in correspondence to  the situation
where almost all roots of the saddle-point equations condense to the same,
known, value. In paper \cite{CP-07a} this correspondence has been shown to hold
in the free-fermion case (i.e., when the weights obey $a^2+b^2=c^2$),
where the assumption of condensation allows one to recover
the Arctic Circle (the limit shape of $2$-enumerated ASMs).
Outside the free-fermion case the correspondence between the Arctic
curve and the condensation of roots is only conjectural
(hence we use below the  term `condensation hypothesis') and it seems it could hardly
be proven, at least by the methods at our disposal.

The condensation hypothesis  is based  essentially on the following argument. EFP is, by
construction, able to discriminate the spatial transition from a frozen to a temperate region
in the scaling limit (jumping  from one to zero exactly at the Arctic curve, i.e.,
at the curve where the spatial phase transition from order to disorder takes place).
Thus, if the Arctic curve exists, the multiple integral representation for EFP must
exhibit the above mentioned stepwise behaviour. In the free-fermion case, this stepwise behaviour
is explained by the mechanism of condensation of roots of the saddle-point equations, which in turn
can be reconducted   to certain specific
properties  of the multiple integral  representation for EFP.
These properties appear  to hold independently of the values of weights,
thus supporting the validity of the condensation hypothesis beyond the free-fermion case.
This allows us to formulate a recipe for deriving an equation for the Arctic curve.
The total procedure is fulfilled here for the two cases of weights corresponding
to $1$- and $3$-enumeration of  ASMs.

In particular,  the resulting equation for the Arctic curve describing
the limit shape of large ASMs is in good agreement with the most refined numerical
simulations available \cite{W}. This could be regarded as confirming
the condensation hypothesis together with the method used
here for deriving of the Arctic curve.

Our paper is organized as follows. In the next Section we
recall results of \cite{CP-07b} on multiple integral representations.
In Section 3 the condensation hypothesis is discussed, and a
procedure for the derivation of limit shapes is explained.
In Section 4  we derive the limit shape  of $1$- and $3$-enumerated ASMs.

\section{Emptiness formation probability}

As in \cite{CP-07a,CP-07b}, we call emptiness formation probability
(EFP) and denote by $F_N^{(r,s)}$, where $r,s=1,\dots,N$,
the probability of having all
arrows on the first $s$ horizontal edges from the top of the lattice,
located between $r$-th and $(r+1)$-th vertical lines (counted from the
right), to be all pointing left.

With this definition, EFP measures the probability that all vertices in
the top-left $(N-r)\times s$ sublattice have the same configuration of
arrows, namely, with all arrows pointing left or downwards, or,
equivalently, that these vertices are of type two (see
Figure~\ref{fig-vertices}). This follows from the peculiarity of both DWBC
and the six-vertex model rule of two
incoming and two outgoing arrows at each lattice vertex (known also as
the `ice-rule').
It is worth noticing that in principle one could define analogous  correlation functions
for the other three corners of the lattice. They would measure the probability
that all vertices in a  bottom-right (or top-right, or bottom-left) rectangular
sublattice of given size are of type one (or three, or four, respectively). It is clear
that all these  correlation functions can be obtained from the EFP defined
above on the basis of symmetry considerations.

In the language of ASMs the definition of EFP implies
that it measures the probability that all entries in the top-left
$(N-r)\times s$ block are $0$'s.  Clearly, fixing the size of ASMs, $N$,
one can expect that EFP takes values closer to $1$ as $r$ and $s$
approach the top left corner of the matrix, and takes values
closer to $0$ otherwise, e.g., while $r$ and $s$ approach its
central region. More precisely, from the definition of EFP one expects that
it is a non-increasing function of the variables
$N-r$ and $s$ for arbitrary fixed value of $N$.

To address the problem of spatial phase separation in the domain-wall
six-vertex model and, therefore, the problem of the limit shape of ASMs,
one has to consider the so-called scaling limit, i.e., the limit of large $N,r,s$
with the ratios $r/N$ and $s/N$ kept fixed.
Using standard arguments of statistical mechanics,
one can argue that, if spatial phase separation occurs,
then in the scaling limit EFP must be equal to $1$ in the frozen region in the top-left
corner of the lattice and to $0$ in the disordered region in the centre.
In other words, if spatial separation occurs, in the scaling limit EFP
is expected  to exhibit stepwise behaviour, with the jump occurring exactly
at  (the top-left `portion' of) the Arctic curve.

In \cite{CP-07b} several equivalent representations for EFP were given.
For what follows we shall need two representations in terms of multiple integrals.
To recall these formulae, we introduce some objects first.

An important role in our considerations is played by the function
$h_N(z)=h_N(z;\Delta,t)$, where $\Delta$ and $t$
are to be regarded as parameters,
\begin{equation}
\Delta:=\frac{a^2+b^2-c^2}{2ab},\qquad t:=\frac{b}{a},
\end{equation}
while $z$ is to be treated as a variable.
This function, defined as a generating function, is
a polynomial of degree $(N-1)$ in $z$,
\begin{equation}
h_N(z):=\sum_{r=1}^{N}H_N^{(r)}z^{r-1},\qquad h_N(1)=1.
\end{equation}
Here the quantity $H_N^{(r)}=H_N^{(r)}(\Delta,t)$
is a boundary correlation function, introduced in \cite{BPZ-02}.
Namely, it is the probability that the sole
vertex of type six (having weight $c$), residing in the first row, appears
at $r$-th position from the right.
At $t=1$ the function $h_N(z)$ has a special meaning as the
generating function for refined $q$-enumerations of ASMs, with
$q$ and $\Delta$ related by $\Delta=1-q/2$. Our derivation
of the limit shapes below involves essentially the known explicit expressions
for this generating function at some particular values of $q$.

For $s=1,\dots,N$, we define functions
\begin{equation}\label{hNs}
h_{N,s}(z_1,\dots,z_s) =
\prod_{1\leq j<k \leq s}^{}(z_j-z_k)^{-1}
\det_{1\leq j,k \leq s} \left[z_j^{k-1}(z_j-1)^{s-k}  h_{N-k+1}(z_j)\right].
\end{equation}
These functions can
be regarded as multi-variable generalizations of $h_N(z)$, in the sense that
$h_{N,1}(z)=h_N(z)$.
It can be easily checked that
\begin{equation}\label{at1}
h_{N,s+1}(z_1,\dots,z_s,1)=h_{N,s}(z_1,\dots,z_s).
\end{equation}
One also has
\begin{equation}\label{at0}
h_{N,s+1}(z_1,\dots,z_s,0)=h_N(0)h_{N-1,s}(z_1,\dots,z_s).
\end{equation}
Properties \eqref{at1} and \eqref{at0} are used in what follows.

We are now ready to turn to the multiple integral representations.
In \cite{CP-07b},
the following multiple integral representation has been obtained
\begin{multline}\label{MIR1}
F_N^{(r,s)} = \frac{(-1)^s}{(2\pi \rmi)^s}
\oint_{C_0}^{} \cdots \oint_{C_0}^{}
\prod_{j=1}^{s}\frac{[(t^2-2t\Delta)z_j+1]^{s-j}}{z_j^r(z_j-1)^{s-j+1}}\,
\\ \times
\prod_{1\leq j<k \leq s}^{} \frac{z_j-z_k}{t^2z_jz_k-2t\Delta z_j+1}\;
h_{N,s}(z_1,\dots,z_s)
\,\rmd z_1\cdots \rmd z_s.
\end{multline}
Here $C_0$ denotes some simple anticlockwise oriented contour
surrounding the point $z=0$ and no other singularity of the integrand.
Formula \eqref{MIR1} has been derived in \cite{CP-07b},
using the quantum inverse scattering method \cite{TF-79,KBI-93}
and the results of papers \cite{I-87,ICK-92,BPZ-02}.
This formula has also been discussed in \cite{CP-07a}
(see equation (3.6) therein).

In \cite{CP-07b}, the following equivalent representation
has been also given:
\begin{multline}\label{MIR2}
F_N^{(r,s)}=\frac{(-1)^{s(s+1)/2}Z_s}{s!(2\pi\rmi)^s a^{s(s-1)}c^s}
\oint_{C_0}^{} \cdots \oint_{C_0}^{}
\prod_{j=1}^{s} \frac{[(t^2-2t\Delta)z_j+1]^{s-1}}{z_j^r(z_j-1)^s}\,
\\ \times
\prod_{\substack{j,k=1\\ j\ne k}}^{s} \frac{1}{t^2 z_jz_k-2t\Delta z_j +1}
\prod_{1\leq j<k\leq s}^{} (z_k-z_j)^2
\\ \times
h_{N,s}(z_1,\dots,z_s)h_{s,s}(u_1,\dots,u_s)
\,\rmd z_1\cdots \rmd z_s.
\end{multline}
Here $Z_s$ denotes the partition function of the six-vertex model
with DWBC on an $s\times s$ lattice, and
\begin{equation}\label{uofz}
u_j:=-\frac{z_j-1}{(t^2-2t\Delta)z_j+1}.
\end{equation}
For later use we emphasize that the integrand of \eqref{MIR2}
is just the symmetrized version of the integrand of \eqref{MIR1}, with
respect to permutations of the integration variables $z_1,\dots,z_s$.
This follows through the
symmetrization procedure explained in \cite{KMST-02}, and some additional
identity proven in \cite{CP-07b}.

We are interested in the behaviour of EFP in the so-called
scaling limit, that is in the limit where $r$, $s$ and  $N$  are all
large, with the ratios $r/N$ and $s/N$ kept finite (and smaller than $1$).
Applying  standard arguments of  saddle-point analysis to representation
\eqref{MIR2}, namely by writing its integrand in exponential form, and by setting
the partial derivatives  of the exponent equal to zero,
we obtain the following  system of coupled saddle-point
equations
\begin{multline}\label{SPE}
-\frac{s}{z_j-1}+ \frac{s(t^2-2t\Delta)}{(t^2-2t\Delta)z_j+1}
-\frac{r}{z_j}
+\sum_{\substack{k=1\\ k\ne j}}^{s}
\frac{2}{z_j-z_k}
\\
+\sum_{\substack{k=1\\ k\ne j}}^{s}
\left(\frac{2\Delta t- t^2 z_k}{t^2z_jz_k-2\Delta t z_j +1}-
\frac{t^2 z_k}{t^2z_jz_k-2\Delta t z_k +1}\right)
+\frac{\partial}{\partial z_j} \log h_{N,s}(z_1,\dots,z_s)
\\
-\frac{t^2-2\Delta t+1}{[(t^2-2\Delta t)z_j+1]^2}
\frac{\partial}{\partial u_j} \log h_{s,s}(u_1,\dots,u_s)
=0.
\end{multline}
In deriving these equations we have used the fact that quantities like
$\log h_{N,s}$ are of order $s^2$, and that
their derivatives with respect to $z_j$'s are of order $s$;
all sub-leading contributions (estimated as $o(s)$) are neglected.

To evaluate the asymptotic behaviour of EFP in the scaling limit, which
would allow one to address the problems of existence of spatial phase
separation and find the corresponding Arctic curve, one needs, in
principle, to be able to describe the solutions of saddle-point equations
\eqref{SPE}. Apparently, this task is a formidable one, at least because
the last two terms in equations \eqref{SPE} are rather implicit. Even in the
technically simpler case of $\Delta=0$ (the free-fermion point) when
these two terms can be found explicitly, the task remains rather
complicated since in this case the saddle-point equations correspond to a
matrix model with a triple logarithmic singularity, or a `triple' Penner
model, and finding their solutions in general settings represents an open
problem (see, for further details, discussion in \cite{CP-07a}).

Nevertheless, it turns out that some interesting information can be
extracted from saddle-point equations \eqref{SPE} provided we assume that
some facts hold true a priori. Namely, we shall assume that EFP in the
scaling limit develops a stepwise behaviour, with the jump from one to
zero occurring at (the top-left portion of) the Arctic curve.
In other words, we assume that the phase separation, and hence the
Arctic curve, exists.

The existence of phase separation can be argued from statistical mechanics,
by studying, for example, the  influence of boundary conditions on the free
energy per site \cite{KZj-00}. Phase separation  is
also supported by previous studies  on  the six-vertex model from which we
know that usually the phenomena are qualitatively similar for the whole
range of values of the weights corresponding to the same regime (in the
phase diagram) of the model \cite{B-82}. On the other hand, from domino tilings
studies it is known that the phase separation occurs in  free-fermion
case, $\Delta=0$. Hence we can expect that this is a general fact for
the whole disordered regime, $|\Delta|<1$; the validity of this statement
is supported by numerics \cite{SZ-04,AR-05}.

Assuming a priori the existence of the limit
shape, or equivalently, the stepwise behaviour of EFP in the scaling limit,
we shall argue that the location of the step correspond to a very particular,
and relatively simple solution of the saddle-point equations \eqref{SPE}.
This, in turn, will gives us a recipe for obtaining the  equation of the  Arctic curve,
and hence the limit shape of ASMs.

\section{Condensation hypothesis and `reduced saddle-point equation'}

As already discussed, by assuming the existence of the phase separation,
we require that EFP in the scaling limit, $N,r,s\to \infty$ with $r/N$
and $s/N$ kept fixed, has a stepwise behaviour, with the jump at
the Arctic curve. Hence to address the problem of finding this curve
one can try to explain how the multiple integral for EFP may have such a
behaviour in the limit.

Intuitively speaking, it is clear that the value of EFP in the ordered
region, where it must be equal to $1$, can be explained by the fact that
in this case the roots of equations \eqref{SPE} are such that integral
\eqref{MIR2} is governed by residues from certain poles rather than by
contributions from the corresponding saddle-point contours. Below we show
that these are the poles (for each integration variable) at point $z=1$,
and prove that indeed the cumulative residue at these poles over all
variables is exactly $1$. On the other hand, the value of
EFP in the disordered region, where it must be equal to $0$, can be
explained simply by the fact that in this case no pole contributes to
the integral \eqref{MIR2}, which is thus given by integration over the
saddle-point contours, vanishing in the scaling limit.

According to the above picture the  transition from one region to another,
i.e., when EFP jumps from $1$ to $0$, can be associated to the situation
where almost all the roots of saddle-point equations are located
at the point $z=1$, i.e., in correspondence to the pole relevant for the
stepwise behaviour of EFP.  Here `almost all' means all but a vanishing fraction of roots.
This is a very special solution of saddle-point equations,  characterized
by  what can be called  `total condensation' of roots, which is defined more
precisely below.  It turns out that the condition of total condensation can be implemented
efficiently, leading to the expression for a curve in the unit square of the scaling limit variables,
which, in view of the above considerations, we recognize as the Arctic curve.

In particular, the validity of the assumption of total condensation can be verified
in the free-fermion case, $\Delta=0$. In \cite{CP-07a}, it has been shown
that indeed in this case the limit shape corresponds to the
situation where almost all roots of equations \eqref{SPE}, as $s\to\infty$,
condense at the point $z=1$. More precisely, it has been shown that
assuming total condensation one can recover
the Arctic curve, which in this case is the Arctic Circle (or ellipse),
already known from previous studies \cite{JPS-98}. On the other hand,
as it can be easily seen from the equations in \cite{CP-07a},
restricting to the Arctic circle, one obtains for
the resolvent of the saddle-point solutions a trivial expression,
having just a single pole at $z=1$, which implies precisely total condensation
(see also discussion below).
Thus, at $\Delta=0$ one can verify the full correspondence between total
condensation and the Arctic curve.

The possibility of total condensation for random matrix integrals in general,
and for  multiple integral \eqref{MIR2} when specialized to the case of $\Delta=0$ in particular,
can be related to a simple fundamental property, namely, to the presence in the integrand of a
pole at the same point in all $s$ integration variables, $z_1,\dots,z_s$, which moreover
must be exactly of order $s$ (see discussion in \cite{PW-95,AKM-94}).
On the other hand, in the case of $\Delta=0$, the unit jump in the stepwise behaviour
of the multiple is due to the fact that the cumulative residue over all
variables $z_1,\dots,z_s$ at this pole is equal to $1$, i.e., to
the value of the jump.

It is a remarkable fact that for the multiple integral representation for EFP
these two crucial properties hold for generic values of $\Delta$.
The first property can be easily verified by direct inspection of formula
\eqref{MIR2}. Verification of the second property could represent a
difficult problem, but fortunately it can easily solved provided
that multiple integral representation \eqref{MIR1} rather than
\eqref{MIR2} is used for the analysis. Indeed, basing on multiple
integral representation \eqref{MIR1}, let us consider, for
$r,s=1,\dots,N$, the integral
\begin{multline}\label{Irs}
I_N^{(r,s)} :
= \frac{(-1)^s}{(2\pi \rmi)^s}
\oint_{C_1^-}^{} \cdots \oint_{C_1^-}^{}
\prod_{j=1}^{s}\frac{[(t^2-2t\Delta)z_j+1]^{s-j}}{z_j^r(z_j-1)^{s-j+1}}\,
\\ \times
\prod_{1\leq j<k \leq s}^{} \frac{z_j-z_k}{t^2z_jz_k-2t\Delta z_j+1}\;
h_{N,s}(z_1,\dots,z_s)
\,\rmd z_1\cdots \rmd z_s.
\end{multline}
Here $C_1^-$ is a closed contour in the complex plane enclosing point
$z=1$ and no other singularity of the integrand; the minus in the
notation indicates negative (clockwise) orientation. We have
\begin{equation}\label{I=1}
I_N^{(r,s)}=1.
\end{equation}
Indeed, performing integration in the variable $z_s$, and taking into account
\eqref{at1}, identity \eqref{I=1} follows immediately by induction.
Note that relation \eqref{I=1} implies the same relation
for the integral with the  integrand of \eqref{MIR2}, which
is just the symmetrized version of the integrand of \eqref{MIR1}.

The fact that these two crucial properties still hold for generic values of $\Delta$,
together with the assumed stepwise behaviour of EFP in the scaling limit
lead us to conjecture that the correspondence of  limit shape  and
condensation of (almost) all roots of saddle-point equations \eqref{SPE}
holds for generic values of $\Delta$. Since we are not able to prove
this statement, we call it  here `condensation hypothesis'.

To treat the consequences of condensation of roots,
we rely on results of papers \cite{PW-95,AKM-94}, where an analytic description of
the mechanism of condensation has been provided,
in the context of random matrix models with generic polynomial potentials with an additional
logarithmic term (Penner models, \cite{P-88}). Since these results do not depend on the explicit form
of the potential but only on some  of its properties it is reasonable to assume they
still hold in more general situations, such as the one of multiple integral representation
\eqref{MIR2}, where these properties are again observed, as discussed above.

To consider the scaling limit, $N,r,s\to\infty$, we define variables $x$ and $y$ such that
\begin{equation}
x:=\frac{N-r}{N},\qquad y:= \frac{s}{N},\qquad x,y \in [0,1].
\end{equation}
where, in general, $x,y \in [0,1]$. Specifically, the top-left portion of
limit shape for ASM's will be given as an equation for $x$ and $y$, with
these variables varying in the interval $[0,1/2]$.

By `total condensation' we mean that, as $s\to \infty$,
almost all roots $z_j$ of \eqref{SPE} have the value $z=1$, in the sense
that the amount of non-condensed roots is $o(s)$.
This implies that at $s=\infty$
the density of roots is just the Dirac delta-function $\delta(z-1)$, or that
the resolvent of the solution (the Green function) is $G(z)=(z-1)^{-1}$.
Let $s_c$ and $s_n$ denote the numbers of condensed and non-condensed roots, $s_c+s_n=s$.
Total condensation thus means that $s_c/s\to 1$ and $s_n/s\to 0$, as $s\to \infty$.

For an analytical treatment of the implication of the condensation of roots
in the context of equations \eqref{SPE},  we introduce here the concept of
the `reduced saddle-point equation'. To this purpose
we note that the non-condensed roots are only a vanishing fraction of the total number of roots,
and thus that their contribution into sums over all roots can be neglected as $s$ tends to infinity.
Hence, assuming, without loss of generality,  that $z_j$, where $j=1,\dots,s_n$, is one of the
non-condensed roots, we may pick up the $j$-th equation in \eqref{SPE} and
simply set $z_k=1$ for  $k= s_n+1,\dots,s$. For example, the first sum in \eqref{SPE}
(the fourth term in the first line) then
reduces to $2s/(z_j-1)$, at leading order for large $s$, thus combining with the
first term in \eqref{SPE}. Similarly, one can easily evaluate the second sum,
with a result partially combining with the second term in \eqref{SPE}.

To evaluate contributions from the last two terms in \eqref{SPE} we use
relations \eqref{at1} and \eqref{at0}. For instance,
due to property \eqref{at1} function $h_{N,s}(z_1,\dots,z_s)$ simplifies to function
$h_{N,s_n}(z_1,\dots,z_{s_n})$, which, in turn, for $N,s\to\infty$, and
$s_n/N\sim 0$, can be evaluated for large $s$ directly from its definition
\eqref{hNs}. In this way we obtain
\begin{align}
\log h_{N,s}(z_1,\dots,z_s)\Big|_{z_{s_n+1}=\dots=z_{s}=1}
&=\log h_{N,s_n}(z_1,\dots,z_{s_n})
\notag\\
&=\sum_{k=1}^{s_n} \log h_N(z_k) +o(s).
\end{align}
Here each term under the sum is of order $N$ and hence estimated as of order $s$.
The fact that $\log h_N(z)$ is of order $N$ can be justified using statistical mechanics arguments;
moreover this fact is transparent for the examples considered below.

Similarly, noticing that $u_k\to 0$ as
$z_j\to 1$, see \eqref{uofz}, and using property \eqref{at0}, we find that function
$h_{s,s}(u_1,\dots,u_s)$ simplifies, modulo an unessential factor, to
function $h_{s_n,s_n}(u_1,\dots,u_{s_n})$. Recalling that
$h_{s_n,s_n}(u_1,\dots,u_{s_n})$ is a polynomial of order $s_n$ in each
of its variables, we obtain that its logarithm for large $s$ is estimated
as $o(s)$. Thus we have
\begin{align}
\log h_{s,s}(u_1,\dots,u_{s})\Big|_{z_{s_n+1}=\dots=z_{s}=1}
&= \sum_{k=s_n+1}^{s} \log h_k(0) +\log h_{s_n,s_n}(u_1,\dots,u_{s_n})
\notag\\
&= C_1 s^2  + C_2 s +o(s).
\end{align}
Here  $C_1$ and $C_2$ are some quantities which do not
depend on $u_j$ $(j=1,\dots,s_n)$. After differentiating we find that
the last term  in \eqref{SPE} is estimated as $o(s)$,
and can thus be neglected at the considered order.

As a result, writing simply $z$ for $z_j$,
we arrive at the following equation
\begin{equation}\label{SPEcond}
\frac{y}{z-1}-\frac{1-x}{z}-\frac{t^2 y}{t^2 z-2t\Delta +1}
+ \lim_{N\to\infty} \frac{1}{N}
\frac{\partial}{\partial z} \log h_N(z)=0.
\end{equation}
We call equation \eqref{SPEcond} the reduced saddle-point equation. The
solutions of this equation give the non-condensed roots of \eqref{SPE}
at total condensation.

As observed in \cite{AKM-94}, a necessary condition for the total
condensation of roots of saddle-point equations like
\eqref{SPE}, is the presence of two coinciding
non-condensed roots.
In general this pair of coinciding real roots constitutes
the germ of the cut which the resolvent develops when parameters are varied
with respect to the situation of total condensation.

In our case, in view of the
scenario depicted above,  when the parameters $x,y$ are varied
away from the limit shape, towards the region where EFP assumes the value $1$,
such cut necessarily lies
on the real axis,  with $z>1$. This imply that at condensation, in
correspondence of the limit shape, these coinciding  roots must lie on
the real axis, in the interval $[1,\infty)$.  Their position depends on the
position of point $(x,y)$ on the limit shape. Thus the value of these roots
naturally parameterizes the limit shape between two contact points.

Summarizing, the condensation hypothesis leads
to the following recipe for the derivation of
limit shapes: we require \eqref{SPEcond} to have two coinciding roots,
which moreover must run over the interval $[1,\infty)$. Denoting the
value of these two roots by $\omega$, we shall see below that as $\omega$
runs over the real axis from point $z=1$ to point $z=\infty$, it
parameterizes the top-left portion of the limit shape, from the
contact point on the top (corresponding to $\omega=1$) to the contact point
on the left (corresponding to $\omega=\infty$).

An ingredient which is necessary for this programme, is the knowledge of
the last term in \eqref{SPEcond}. Fortunately, function $h_N(z)$ is
explicitly known in some interesting cases. These are, for $t=1$, the
case of $\Delta=1/2$, corresponding  to usual enumeration of ASMs, and
the case of $\Delta=-1/2$, corresponding to $3$-enumerated  ASMs; for
generic $t$ the function $h_N(z)$ is also known for $\Delta=0$,
corresponding to the six-vertex model in the free-fermion case
(specializing further $t=1$, one gets $2$-enumerated ASMs).

To illustrate how the recipe is working, let us consider the
case of $\Delta=0$. We set also $t=1$ for simplicity,
so that function $h_N(z)=h_N(z;\Delta,t)$ in this case is just
$h_N(z;0,1)=[(z+1)/2]^{N-1}$ (see, e.g., \cite{CP-07a}).
The reduced saddle-point equation reads
\begin{equation}
\frac{y}{z-1} - \frac{1-x}{z}+\frac{1-y}{z+1}=0.
\end{equation}
Denoting by $g(z)$ the function in the left-hand side,
we require $g(z)=(z-\omega)^2\tilde g(z)$ where $\tilde g(z)$ is regular near point $z=\omega$.
This can be implemented by the system of two equations
$g(\omega)=0$ and $g'(\omega)=0$, where the prime denotes derivative,
for unknowns $x$ and $y$. Solving this system, we obtain
\begin{equation}
x=\frac{1}{\omega^2+1},\qquad y=\frac{(\omega-1)^2}{2(\omega^2+1)},\qquad
\omega\in[1,\infty).
\end{equation}
Eliminating $\omega$, we also have the equation for the limit shape
\begin{equation}\label{ASM2eq}
4x(1-x)+4y(1-y)=1.
\end{equation}
Here $x$ and $y$ take values in interval $[0,1/2]$; equation \eqref{ASM2eq}
describes the top-left portion of the Arctic Circle, which is
the limit shape of $2$-enumerated ASMs.

\section{Limit shapes of $1$- and $3$-enumerated ASMs}

We start with considering the case of $1$-enumerated ASMs. In this case
we have $\Delta=1/2$ and $t=1$, and the function
$h_N\big(z;\tfrac{1}{2},1\big)$ is given by the formula (see, e.g.,
\cite{CP-05a,CP-05b})
\begin{equation}
h_N\big(z;\tfrac{1}{2},1\big)=\F{-N+1}{N}{2N}{1-z}.
\end{equation}
We write here the hypergeometric function in such a way that the
third parameter is positive, and larger than the second one, so that
the Euler integral representation for Gauss hypergeometric function
can be used to study the large $N$ limit.

Explicitly, the Euler integral representation gives the following
expression
\begin{equation}\label{euler_rep}
h_N\big(z;\tfrac{1}{2},1\big)=\frac{\Gamma(2N)}{[\Gamma(N)]^2}
\int_0^1 \left[\tau (1-\tau)(1-\tau+z\tau)\right]^{N-1}\,\rmd \tau.
\end{equation}
The large $N$ behaviour of this integral can be found
via the standard saddle-point analysis, which
in fact can provide a uniform asymptotic expression in $z$. Hence
in evaluating the last term in  \eqref{SPEcond} we can use the property that
the logarithmic derivative and the
large $N$ limit commute (this property can also be verified directly
for integral representation \eqref{euler_rep}).
Explicitly, we find that, as $N\to \infty$,
\begin{equation}
\log h_N\big(z;\tfrac{1}{2},1\big)=
N \log\left[4v(1-v)(1-v+zv)\right] + o(N),
\end{equation}
where
\begin{equation}
v:=\frac{2-z-\sqrt{z^2-z+1}}{3(1-z)}.
\end{equation}
Upon differentiating with respect to $z$, we obtain
that the reduced saddle-point equation in this case reads
\begin{equation}\label{rspe-1}
\frac{y}{z-1}-\frac{1-x+y}{z}+\frac{1-\sqrt{z^2-z+1}}{z(1-z)}=0.
\end{equation}

The requirement that equation \eqref{rspe-1} has two coinciding roots
gives the following parametric solution  for the limit shape
of large ASMs:
\begin{equation}\label{ASMxy}
x= 1-\frac{2\omega-1}{2\sqrt{\omega^2-\omega+1}},\qquad
y= 1-\frac{\omega+1}{2\sqrt{\omega^2-\omega+1}},\qquad \omega\in[1,\infty).
\end{equation}
Eliminating $\omega$ from this parametric solution
we obtain that the limit shape is described by the equation
\begin{equation}\label{ASMeq}
4x(1-x)+ 4y(1-y)+ 4xy =1, \qquad
x,y \in \big[0,\tfrac{1}{2}\big].
\end{equation}
Formula \eqref{ASMeq} is the central result of our paper.
A comparison of this formula with available numerical data is discussed below.

Let us now consider the case of $3$-enumerated ASMs
($\Delta=-1/2$ and  $t=1$). The function $h_N\big(z;-\tfrac{1}{2};1\big)$
has been obtained in \cite{CP-05a} (see also \cite{CP-05b}).
The following formulae are valid
\begin{equation}
h_N\big(z;-\tfrac{1}{2};1\big)=
\begin{cases}
(1/2)(z+1) B_{2m}(z) &\text{for}\  N=2m+2, \\
(1/9)(2z+1)(z+2) B_{2m}(z) &\text{for}\  N=2m+3,
\end{cases}
\end{equation}
where $B_{2m}(z)$ is the following polynomial of degree $2m$:
\begin{multline}
B_{2m}(z)=\frac{m+1}{3^{m-1} (2m+3)}\,
z^m (z+2)^m
\F{-m}{m+2}{2m+4}{\frac{z^2-1}{z(z+2)}}
\\
-\frac{m}{3^{m-1} (2m+3)}\,
z^m (z+2)^{m-1}
\F{-m+1}{m+2}{2m+4}{\frac{z^2-1}{z(z+2)}}.
\end{multline}
Again, in comparison with \cite{CP-05b}, we have written
here the hypergeometric polynomials in such a way that the Euler integral formula
can be directly applied.

Evaluating the integrals
through the saddle-point method, we obtain that, as $N\to\infty$,
\begin{equation}
\log h_N\big(z;-\tfrac{1}{2},1\big)=
N \log\left[\frac{2(2z+1)(z+2)}{9(z+1)}\right] + o(N).
\end{equation}
The reduced saddle-point equation reads:
\begin{equation}\label{rspe-3}
\frac{y}{z-1}-
\frac{1-x}{z}-\frac{y}{2+z}+\frac{2z^2+4z+3}{(1+z)
(2+z)(1+2z)}=0.
\end{equation}

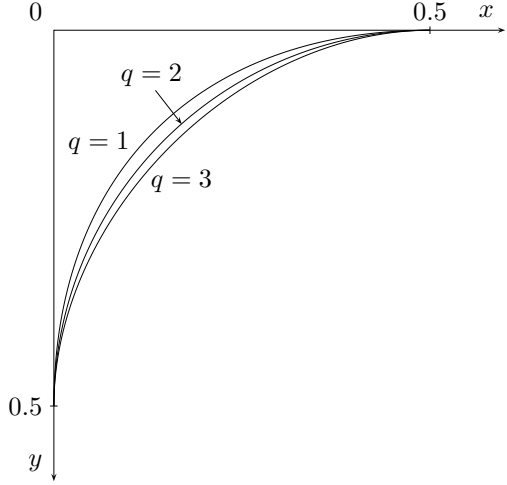
\begin{figure}
\centering

\begin{pspicture}(-.5,.5)(6,-6)
\psset{xunit=10,yunit=10,linewidth=.01}
\savedata{\one}[{{0.5, 0.}, {0.490959, -0.0000548306},
{0.481973, -0.000219317}, {0.473044, -0.00049344}, {0.464173, -0.00087717},
{0.455361, -0.00137047}, {0.446608, -0.00197327}, {0.437917, -0.00268552},
{0.429286, -0.00350714}, {0.420719, -0.00443804}, {0.412215, -0.0054781},
{0.403775, -0.00662723}, {0.395401, -0.0078853}, {0.387093, -0.00925216},
{0.378852, -0.0107277}, {0.37068, -0.0123117}, {0.362576, -0.014004},
{0.354542, -0.0158044}, {0.346579, -0.0177127}, {0.338688, -0.0197288},
{0.330869, -0.0218524}, {0.323124, -0.0240832}, {0.315453, -0.0264211},
{0.307857, -0.0288657}, {0.300337, -0.0314168}, {0.292893, -0.0340742},
{0.285527, -0.0368374}, {0.27824, -0.0397063}, {0.271031, -0.0426805},
{0.263903, -0.0457597}, {0.256855, -0.0489435}, {0.249889, -0.0522316},
{0.243005, -0.0556236}, {0.236204, -0.0591192}, {0.229487, -0.062718},
{0.222854, -0.0664196}, {0.216307, -0.0702235}, {0.209845, -0.0741294},
{0.20347, -0.0781368}, {0.197183, -0.0822454}, {0.190983, -0.0864545},
{0.184872, -0.0907639}, {0.178851, -0.0951729}, {0.172919, -0.0996812},
{0.167079, -0.104288}, {0.161329, -0.108993}, {0.155672, -0.113796},
{0.150107, -0.118697}, {0.144636, -0.123693}, {0.139258, -0.128786},
{0.133975, -0.133975}, {0.128786, -0.139258}, {0.123693, -0.144636},
{0.118697, -0.150107}, {0.113796, -0.155672}, {0.108993, -0.161329},
{0.104288, -0.167079}, {0.0996812, -0.172919}, {0.0951729, -0.178851},
{0.0907639, -0.184872}, {0.0864545, -0.190983}, {0.0822454, -0.197183},
{0.0781368, -0.20347}, {0.0741294, -0.209845}, {0.0702235, -0.216307},
{0.0664196, -0.222854}, {0.062718, -0.229487}, {0.0591192, -0.236204},
{0.0556236, -0.243005}, {0.0522316, -0.249889}, {0.0489435, -0.256855},
{0.0457597, -0.263903}, {0.0426805, -0.271031}, {0.0397063, -0.27824},
{0.0368374, -0.285527}, {0.0340742, -0.292893}, {0.0314168, -0.300337},
{0.0288657, -0.307857}, {0.0264211, -0.315453}, {0.0240832, -0.323124},
{0.0218524, -0.330869}, {0.0197288, -0.338688}, {0.0177127, -0.346579},
{0.0158044, -0.354542}, {0.014004, -0.362576}, {0.0123117, -0.37068},
{0.0107277, -0.378852}, {0.00925216, -0.387093}, {0.0078853, -0.395401},
{0.00662723, -0.403775}, {0.0054781, -0.412215}, {0.00443804, -0.420719},
{0.00350714, -0.429286}, {0.00268552, -0.437917}, {0.00197327, -0.446608},
{0.00137047, -0.455361}, {0.00087717, -0.464173}, {0.00049344, -0.473044},
{0.000219317, -0.481973}, {0.0000548306, -0.490959}, {0., -0.5}}]
\savedata{\two}[ {{0.5, 0.}, {0.492146, -0.0000616838},
{0.484295, -0.00024672}, {0.476447, -0.000555063}, {0.468605, -0.000986636},
{0.46077, -0.00154133}, {0.452946, -0.00221902}, {0.445133, -0.00301952},
{0.437333, -0.00394265}, {0.429549, -0.00498817}, {0.421783, -0.00615583},
{0.414035, -0.00744534}, {0.406309, -0.00885637}, {0.398606, -0.0103886},
{0.390928, -0.0120416}, {0.383277, -0.013815}, {0.375655, -0.0157084},
{0.368063, -0.0177213}, {0.360504, -0.0198532}, {0.35298, -0.0221035},
{0.345492, -0.0244717}, {0.338041, -0.0269573}, {0.330631, -0.0295596},
{0.323263, -0.032278}, {0.315938, -0.0351118}, {0.308658, -0.0380602},
{0.301426, -0.0411227}, {0.294243, -0.0442984}, {0.28711, -0.0475865},
{0.28003, -0.0509862}, {0.273005, -0.0544967}, {0.266035, -0.0581172},
{0.259123, -0.0618467}, {0.252271, -0.0656842}, {0.245479, -0.069629},
{0.238751, -0.0736799}, {0.232087, -0.077836}, {0.225489, -0.0820963},
{0.218958, -0.0864597}, {0.212497, -0.0909251}, {0.206107, -0.0954915},
{0.19979, -0.100158}, {0.193546, -0.104922}, {0.187379, -0.109785},
{0.181288, -0.114743}, {0.175276, -0.119797}, {0.169344, -0.124944},
{0.163494, -0.130184}, {0.157726, -0.135516}, {0.152044, -0.140937},
{0.146447, -0.146447}, {0.140937, -0.152044}, {0.135516, -0.157726},
{0.130184, -0.163494}, {0.124944, -0.169344}, {0.119797, -0.175276},
{0.114743, -0.181288}, {0.109785, -0.187379}, {0.104922, -0.193546},
{0.100158, -0.19979}, {0.0954915, -0.206107}, {0.0909251, -0.212497},
{0.0864597, -0.218958}, {0.0820963, -0.225489}, {0.077836, -0.232087},
{0.0736799, -0.238751}, {0.069629, -0.245479}, {0.0656842, -0.252271},
{0.0618467, -0.259123}, {0.0581172, -0.266035}, {0.0544967, -0.273005},
{0.0509862, -0.28003}, {0.0475865, -0.28711}, {0.0442984, -0.294243},
{0.0411227, -0.301426}, {0.0380602, -0.308658}, {0.0351118, -0.315938},
{0.032278, -0.323263}, {0.0295596, -0.330631}, {0.0269573, -0.338041},
{0.0244717, -0.345492}, {0.0221035, -0.35298}, {0.0198532, -0.360504},
{0.0177213, -0.368063}, {0.0157084, -0.375655}, {0.013815, -0.383277},
{0.0120416, -0.390928}, {0.0103886, -0.398606}, {0.00885637, -0.406309},
{0.00744534, -0.414035}, {0.00615583, -0.421783}, {0.00498817, -0.429549},
{0.00394265, -0.437333}, {0.00301952, -0.445133}, {0.00221902, -0.452946},
{0.00154133, -0.46077}, {0.000986636, -0.468605}, {0.000555063, -0.476447},
{0.00024672, -0.484295}, {0.0000616838, -0.492146}, {0., -0.5}}]
\savedata{\three}[{{0.5, 0.}, {0.492936, -0.0000639701},
{0.485851, -0.000255886}, {0.478747, -0.000575761}, {0.471625, -0.00102361},
{0.464486, -0.00159947}, {0.457331, -0.00230334}, {0.450161, -0.00313524},
{0.442978, -0.00409518}, {0.435783, -0.00518315}, {0.428577, -0.00639913},
{0.421362, -0.00774308}, {0.41414, -0.00921494}, {0.406912, -0.0108146},
{0.39968, -0.012542}, {0.392445, -0.0143969}, {0.385209, -0.0163792},
{0.377975, -0.0184887}, {0.370743, -0.020725}, {0.363517, -0.0230879},
{0.356298, -0.025577}, {0.349088, -0.028192}, {0.341889, -0.0309323},
{0.334703, -0.0337974}, {0.327534, -0.0367868}, {0.320382, -0.0398999},
{0.31325, -0.043136}, {0.306141, -0.0464942}, {0.299058, -0.0499739},
{0.292001, -0.053574}, {0.284975, -0.0572937}, {0.277981, -0.0611319},
{0.271022, -0.0650876}, {0.2641, -0.0691595}, {0.257219, -0.0733465},
{0.25038, -0.0776472}, {0.243587, -0.0820602}, {0.236841, -0.0865842},
{0.230146, -0.0912175}, {0.223505, -0.0959586}, {0.216919, -0.100806},
{0.210391, -0.105757}, {0.203925, -0.110812}, {0.197522, -0.115966},
{0.191186, -0.12122}, {0.184918, -0.126571}, {0.178722, -0.132016},
{0.1726, -0.137554}, {0.166554, -0.143182}, {0.160586, -0.148898},
{0.154701, -0.154701}, {0.148898, -0.160586}, {0.143182, -0.166554},
{0.137554, -0.1726}, {0.132016, -0.178722}, {0.126571, -0.184918},
{0.12122, -0.191186}, {0.115966, -0.197522}, {0.110812, -0.203925},
{0.105757, -0.210391}, {0.100806, -0.216919}, {0.0959586, -0.223505},
{0.0912175, -0.230146}, {0.0865842, -0.236841}, {0.0820602, -0.243587},
{0.0776472, -0.25038}, {0.0733465, -0.257219}, {0.0691595, -0.2641},
{0.0650876, -0.271022}, {0.0611319, -0.277981}, {0.0572937, -0.284975},
{0.053574, -0.292001}, {0.0499739, -0.299058}, {0.0464942, -0.306141},
{0.043136, -0.31325}, {0.0398999, -0.320382}, {0.0367868, -0.327534},
{0.0337974, -0.334703}, {0.0309323, -0.341889}, {0.028192, -0.349088},
{0.025577, -0.356298}, {0.0230879, -0.363517}, {0.020725, -0.370743},
{0.0184887, -0.377975}, {0.0163792, -0.385209}, {0.0143969, -0.392445},
{0.012542, -0.39968}, {0.0108146, -0.406912}, {0.00921494, -0.41414},
{0.00774308, -0.421362}, {0.00639913, -0.428577}, {0.00518315, -0.435783},
{0.00409518, -0.442978}, {0.00313524, -0.450161}, {0.00230334, -0.457331},
{0.00159947, -0.464486}, {0.00102361, -0.471625}, {0.000575761, -0.478747},
{0.000255886, -0.485851}, {0.0000639701, -0.492936}, {0., -0.5}}]
\dataplot{\one}
\dataplot{\two}
\dataplot{\three}
\psset{unit=1}
\psline{<->}(0,-6)(0,0)(6,0)
\rput[r](-.15,.25){$0$}
\psline(5,-.05)(5,.05) \rput(5,.25){$0.5$} \rput(5.75,.25){$x$}
\psline(-.05,-5)(.05,-5) \rput[r](-.15,-5){$0.5$} \rput[r](-.15,-5.75){$y$}
\rput(.6,-1.5){$q=1$}
\rput(1.3,-.6){$q=2$} \psline{->}(1.35,-.8)(1.7,-1.25)
\rput(1.7,-2){$q=3$}
\end{pspicture}
\caption{The limit shapes of $q$-enumerated ASMs for $q=1,2,3$, given by
equations \protect\eqref{ASMeq}, \protect\eqref{ASM2eq}, and \protect\eqref{ASM3eq},
respectively.}
\label{fig-limshape}
\end{figure}

The requirement that equation \eqref{rspe-3} has two coinciding roots
gives us the following parametric solution  for the limit shape
of $3$-enumerated ASMs:
\begin{align}\label{ASM3xy}
x&=\dfrac{7\omega^4+14\omega^3+19\omega^2+12\omega+2}
{(\omega^2+2)(2\omega+1)^2(\omega+1)^2},
\notag\\
y&=\dfrac{(\omega-1)^2(6\omega^4+16\omega^3+19\omega^2+16\omega+6)}
{3(\omega^2+2)(2\omega+1)^2(\omega+1)^2},\qquad
\omega\in[1,\infty).
\end{align}
Equivalently, one can search for the equation connecting $x$ and $y$.
In this way we obtain that the limit shape of $3$-enumerated ASMs is described
by the following sextic equation:
\begin{multline}\label{ASM3eq}
324\, x^6 + 1620\, x^5 y + 3429\, x^4 y^2 + 4254\, x^3 y^3
+ 3429\, x^2 y^4 + 1620\, x y^5 + 324\, y^6
\\
- 972\, x^5 - 1458\, x^4 y - 2970\, x^3 y^2 - 2970\, x^2 y^3
- 1458\, x y^4 - 972\, y^5
\\
- 6147\, x^4 - 9150\, x^3 y - 17462\, x^2 y^2 - 9150\, x y^3 - 6147\, y^4
\\
+ 13914\, x^3 + 24086\, x^2 y + 24086\, x y^2 + 13914\, y^3
\\
- 11511\, x^2 - 17258\, x y - 11511\, y^2
\\
+4392\, x + 4392\, y - 648=0, \qquad x,y\in\big[0,\tfrac{1}{2}\big].
\end{multline}
One can verify directly that this equation
is indeed satisfied by $x$ and $y$ given by \eqref{ASM3xy}.

Figure~\ref{fig-limshape} shows plots of the
limit shapes for $1$-, $2$- and $3$-enumerated ASMs.
The area of the temperate region decreases
as $q$ increases, in agreement with both analytical
and numerical considerations \cite{KZj-00,SZ-04,AR-05}.

\begin{figure}
\centering
\includegraphics[width=.8\textwidth]{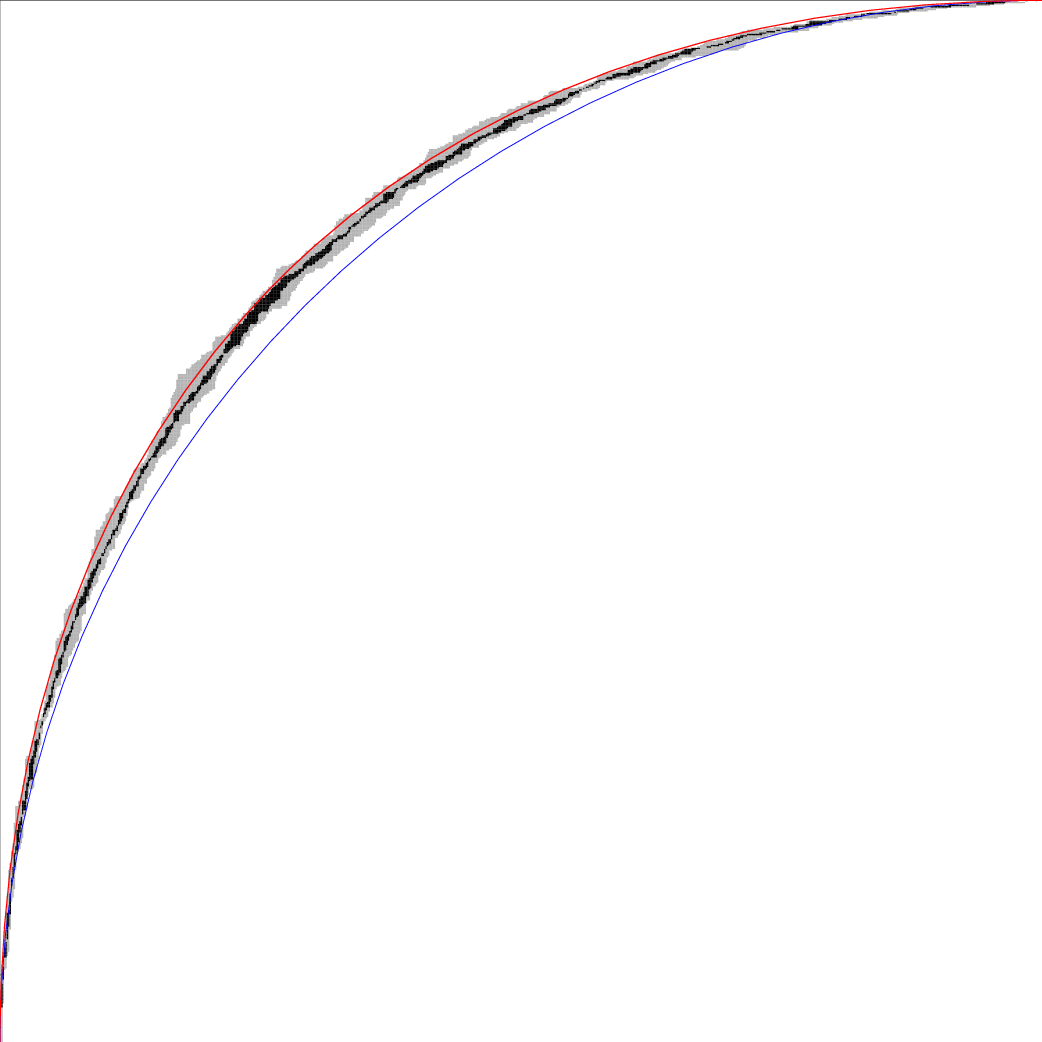}
\caption{Comparison of equation \protect\eqref{ASMeq} for the limit shape of large ASMs,
the outer curve (in red), with the numerical data of \protect\cite{W} for $N=1500$. The inner curve (in blue)
is the Arctic circle,  equation \protect\eqref{ASM2eq}, plotted here for reference.}
\label{fig-numerics}
\end{figure}

Concerning expression \eqref{ASMeq} for the limit shape of  large ASMs, it is
worth mentioning here a comparison of this result with numerical simulations.
The most refined numerical simulations of large ASMs (up to $N=1500$)
presently available have been performed by Wieland \cite{W}, who also
provided pictures comparing these numerical data with equation
\eqref{ASMeq}\footnote{Pictures and the C program code are vailable upon request.}.
As the size of ASMs increases, convergence to this
curve is observed. Although the convergence is rather slow, and more refined
data would be welcome, a good agreement is already obtained.

In figure~\ref{fig-numerics} we have elaborated the data of \cite{W} for ten
random samples of ASMs of size $N=1500$, generated using Propp-Wilson `coupling from the past'
algorithm (see \cite{W} for further details). Each pixel of the picture corresponds to one
of the entries of (the top-left quarter of) the matrix, and has been  assigned to one of
five bins, according to the frequency of being, in the ten random samples of ASMs,
in the frozen region. The  five bins are defined by the breakpoints of
$5\%$, $35\%$, $65\%$,  and $95\%$. The picture plots  the three central bins in grey,
black, and grey again, respectively. The picture plots also the analytic expression
\eqref{ASMeq}  for the limit shape of  large ASMs (the outer curve, in red), and, for reference,
equation \eqref{ASM2eq} for the Arctic Circle  (the inner curve, in blue).

\section*{Acknowledgements}

We thank B.~Wieland for providing us his data on heavy simulations for the limit shape
of ASMs and making  useful pictures of them.
This work is partially done within the European Science Foundation
program INSTANS. One of us (AGP) is supported in part by Russian
Foundation for Basic Research (grant 07-01-00358), and by the programme
``Mathematical Methods in Nonlinear Dynamics'' of Russian Academy of
Sciences. AGP also thanks INFN, Sezione di Firenze, where part of this
work was done, for hospitality and support.
Finally, we would like to thank
two anonymous referees for their constructive comments
which stimulated us to improve the presentation of the paper.



\end{document}